\documentclass[prl,twocolumn,showpacs,preprintnumbers,amsmath,amssymb]{revtex4} 

\usepackage{graphicx}
\usepackage{dcolumn}
\usepackage{bm}
\begin{document}

\title{Inefficient emergent oscillations in intersecting driven many-particle flows}
\author{Rui Jiang, Dirk Helbing, Pradyumn Kumar Shukla}
\affiliation{Dresden University of Technology, Andreas-Schubert-Str. 23, 01062 Dresden, Germany}
\author{Qing-Song Wu}
\affiliation{University of Science and Technology of China, Hefei 230026, P.R.China}
\begin{abstract} 
Oscillatory flow patterns have been observed in many different driven many-particle systems.
The conventional assumption
is that the reason for emergent oscillations in opposing flows is an increased efficiency (throughput).
In this contribution, however, we will study intersecting pedestrian and vehicle flows as
an example for inefficient emergent oscillations. In the
coupled vehicle-pedestrian delay problem, oscillating
pedestrian and vehicle flows form when pedestrians cross the street with a small time
gap to approaching cars, while both pedestrians and vehicles benefit, when
they keep some overcritical time gap. That is, when the safety time gap of pedestrians is
increased, the average delay time of pedestrians decreases and 
the vehicle flow goes up. This may be interpreted as a slower-is-faster effect. The 
underlying mechanism of this effect is explained in detail. 
\end{abstract}
\pacs{89.40.+k,
47.54.+r,
89.75.Kd,
47.62.+q}
\maketitle

Emergent oscillations have been discovered in so different systems as the
density oscillator \cite{Steinbock}, ticking hour glass \cite{ticking}, RNA Polymerase traffic on DNA
\cite{Sneppen}, pedestrians passing a bottleneck \cite{ped,panic}, or 
ants \cite{Dussu}. Despite the different underlying oscillation mechanisms and system components,
all of these systems can be treated as driven or self-driven many-particle systems \cite{Helbing}
characterized by counterflows.
Therefore, one might assume a unified principle behind these emergent oscillations such
as an optimization of throughput: The clustering of units with the same flow direction could reduce
``frictional'' interactions, which are particularly high between units moving in different directions. 
In this contribution, we will study the example of intersecting pedestrian and vehicle flows. 
This system is found to show a transition to emergent oscillations as well. However, contrary to
our expectations, the oscillations are not an efficient pattern of motion. Instead, they are related with 
a considerable reduction of the throughput and increased waiting times. 
\par
In the past, 
the investigation of vehicle and pedestrian streams by means of experiments and models from
statistical physics or fluid-dynamics has revealed the mechanisms behind many observed phenomena 
such as different forms of congestion \cite{Helbing,reviews,letters}. Moreover, various
self-organization phenomena \cite{ped,panic} have been discovered in pedestrian flows, including 
the so-called ``faster-is-slower effect'' in ``panicking'' crowds \cite{panic}. These
have stimulated research in many other fields such as colloidal \cite{colloid} and
biological \cite{biology} systems.
\par
Let us now come back to the problem of interacting vehicle and pedestrian flows, a problem that
has not been thoroughly studied in the past. In a way, the problem can be viewed as two dynamically
coupled queues, which cannot be served simultaneously, since pedestrians must cross the street at times when
no vehicle passes and vice versa. Such coupled queuing systems are known to display interesting dynamic
behaviors, including irregular oscillations \cite{Witt} and chaos \cite{chaos}. We
are, therefore, interested in identifying the possible dynamic behaviors of the coupled vehicle-pedestrian-delay
problem, their performance and preconditions.
\par
The pedestrian delay problem is a growing concern of urban planning.
It is defined as follows: Suppose there is a stream of vehicle traffic moving on a main street, and suppose
that a pedestrian arrives at time $t=t_0$ at the roadside and intends to
cross this street (away from any pedestrian crossing facility), see Fig.~1(a), (b). 
What is then the average delay to the pedestrian?
The early pedestrian delay models assume that there is a negative exponential distribution of vehicular headways
\cite{1,2}.
Other models have adopted a shifted exponential distribution, a double-displaced negative 
exponential distribution, etc.\cite{3}.
Recently Guo et al. \cite{6} have proposed a pedestrian delay model, in which the overall delay to pedestrians
is obtained as a combination of the delay by traffic-light induced vehicle clusters
and the delay to pedestrians arriving during the random vehicle flow between the clusters.

Note that, in the pedestrian delay problem, the interactions between vehicles and pedestrians, 
at least influences of pedestrians on vehicles, have not been considered, yet. 
Usually, it is assumed that the crossing of pedestrians will not affect the motion of vehicles. 
This is certainly not realistic. 
Therefore, this paper studies the coupled vehicle-pedestrian delay problem, taking into account 
mutual interactions. This is relevant for the capacity
of traffic infrastructures for both, vehicles and pedestrians.

Our delay model is as follows: Firstly, within one incremental time step $dt$ of 0.1s 
(corresponding to the applied time discretization of the car-following model), 
we assume the arrival of one pedestrian along the roadside with probability $p$. 
When a pedestrian arrives along the roadside at a given point O, he or she checks the traffic
situation [Fig.1(a)]. We will distinguish two situations:
(i) No other pedestrian is on the road. In this case we suppose that, when the safety criterion
\begin{equation}
 d > d_0 + \sigma t_{\rm r} v_n 
\label{safety}
\end{equation}
is satisfied, the pedestrian will cross the road. 
Here, $d_0$ is the minimum safety distance of pedestrians, 
$t_{\rm r}$ the time needed for a pedestrian to traverse a one-lane street, 
$\sigma$ a safety coefficient, $d$ the distance 
from the nearest vehicle $n$ upstream of point O, and $v_n$ its velocity \cite{footnote1}.
(ii) Other pedestrians are crossing the road. In this case, the pedestrians on the road will encourage
newly arriving pedestrians to follow, as an obstructed driver-vehicle unit would not dare to accelerate.
This effect can be simulated by adjusting the safety coefficient $\sigma$. 

We assume that in case (i), the safety coefficient $\sigma$ chosen by a pedestrian is $\sigma_0$, 
while in case (ii), he or she will choose a smaller safety coefficient $\sigma_1$.
The pedestrian simulation is similar to that described in Ref. \cite{7}:
First, the position $x_n(t)$ of the nearest vehicle $n$ upstream of the crossing point O is identified. 
If a pedestrian is on the street, the net distance to the next object is specified as 
$\Delta x(t) = d(t) =x_O-x_n(t)$, and the velocity of the object ahead into the driving direction is $v=0$. 
Otherwise the distance and velocity are given by the next vehicle $n-1$ ahead, i.e. $\Delta x(t)=x_{n-1}(t)-x_n-l_{n-1}(t)$
and $v(t)=v_{n-1}(t)$. This enters the equation of vehicle motion
\begin{equation}\label{e1}
\frac{dv_n}{dt}=f(\Delta x, v, v_n)+\xi_n(t) \, .
\end{equation}
Here, $l$ is the vehicle length, $f$ the acceleration function and $\xi$ a stochastic term.
For illustrative purposes, the acceleration function has been specified according to 
the well-investigated intelligent driver model (IDM) \cite{8}:
\begin{equation}\label{e2}
f(\Delta x, v, v_n)=a\left[1-\left(\frac{v_n}{v_0}\right)^4-\left(\frac{s^*}{\Delta x}\right)^2\right] \, .
\end{equation}
The parameter $v_0$ denotes the desired velocity, while $s^*=s_0+Tv_n+\frac{v_n(v_n-v)}{2\sqrt{ab}}$
is the desired minimum gap,  where $s_0$ is the minimum safety distance of cars, $T$ is the safe time gap, 
$a$ the maximum acceleration, and $b$ the desired deceleration.
The stochastic term $\xi$ has been set to zero.
\begin{figure}
\begin{center}
\hspace*{-2mm}\includegraphics[width=4.3cm]{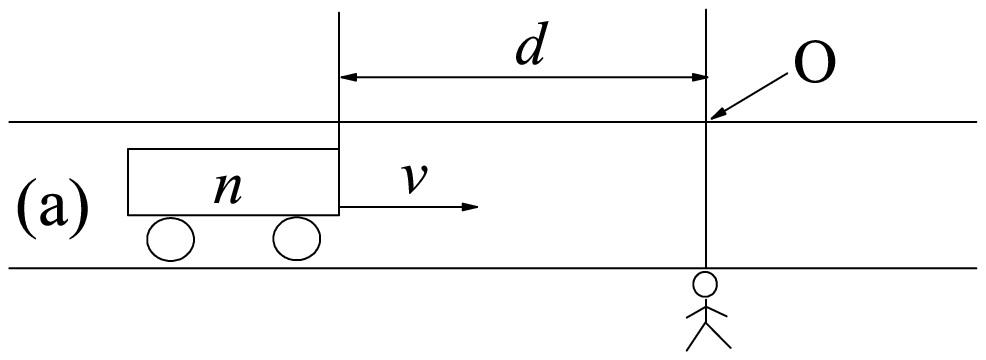}
\includegraphics[width=4.3cm]{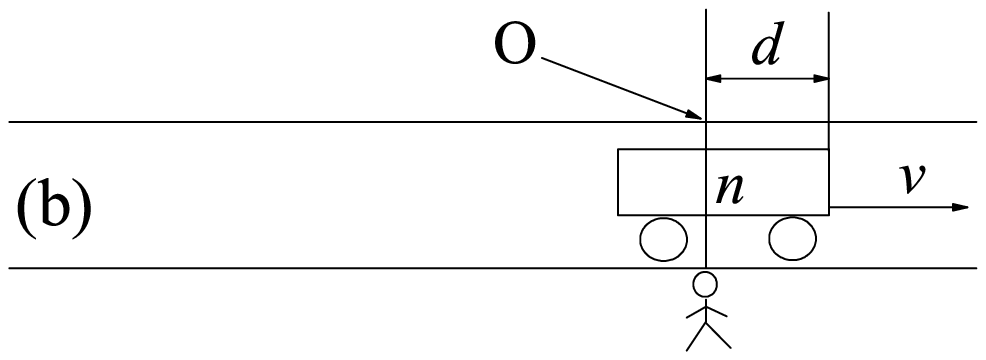} \\[-4mm]
\includegraphics[width=8.8cm]{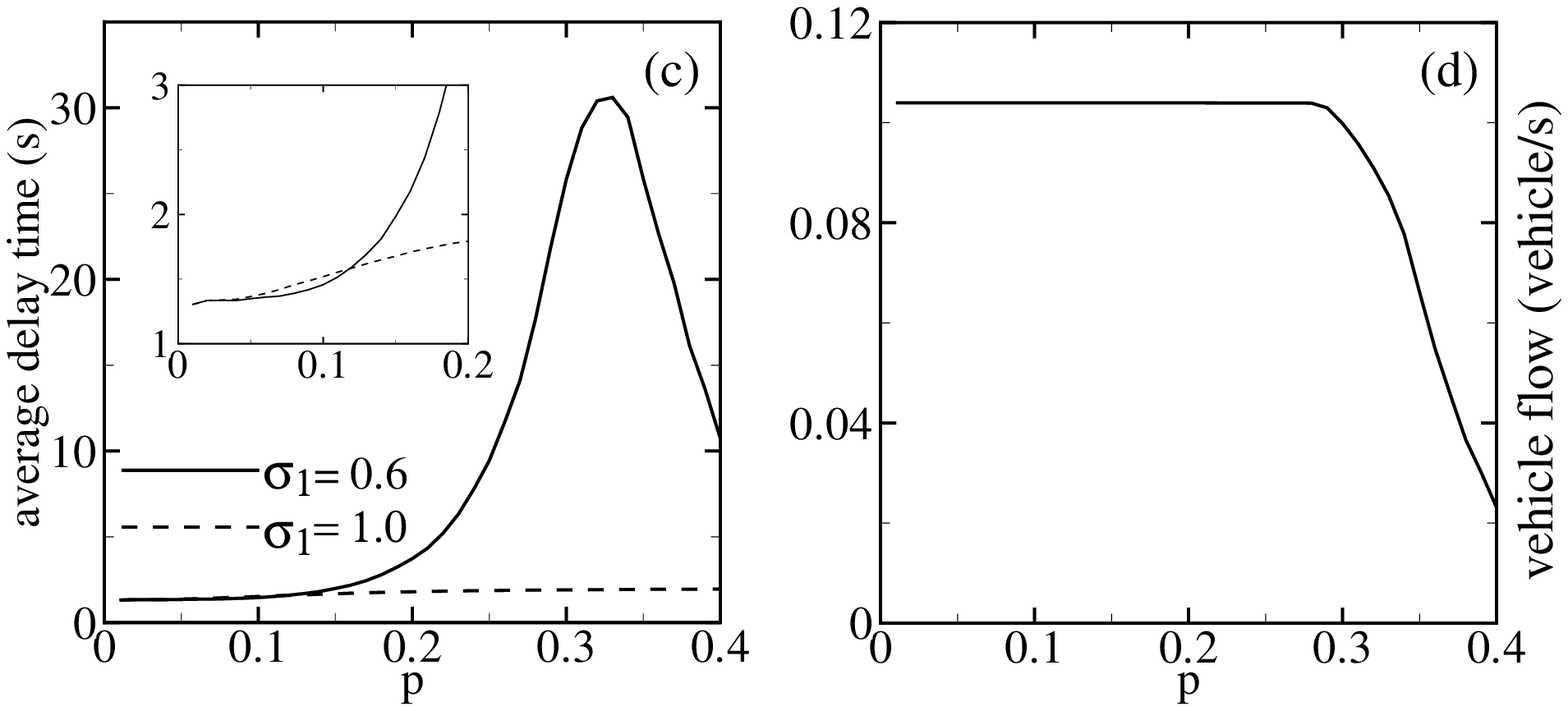}
\vspace*{-11mm}
\end{center}
\caption{(a), (b) Sketch of the vehicle-pedestrian delay system.
(c) Average delay time of pedestrians and (d)
average vehicle flow for $\sigma_1 = 0.6$, both as a function of the pedestrian arrival probability $p$.}
\end{figure}

In each simulation time step, if a random number $y$ uniformly distributed between 0 and 1 is smaller than $p$, 
a new pedestrian occurs at point O. 
The pedestrian crosses the road if the safety criterion (\ref{safety}) is satisfied. 
Otherwise, he or she will wait until next time step.

In the simulations, we adopt open boundary conditions. 
Suppose that a car has been added at time $t=t_0$, then another car is added at place $x=0$ and 
time $t=t_0+ \Delta t$, provided that the previous car is at least 7 m away 
\cite{footnote3}. Here 
$\Delta t$ is determined by a shifted exponential distribution:
$\Delta t=-\frac{ \ln(1-z)}{0.13 \;{\rm s}^{-1}}+2\;\mbox{s}$,
where $z$ is a uniformly distributed random number between 0 and 1. For simplicity,
the speed of the new car is assumed to agree with the one of the car ahead. 
At the exit ($x=L$), cars are removed. If there is no leading car, the distance $\Delta x$ is set to some large number 
and $v$ to the maximum velocity $v_0$.

In the simulations,  Eq.~(\ref{e1}) is solved by the Euler method. The time step is set to 0.1 s,
as smaller values do not change our results.
The model parameters are $v_0=15 $ m/s, $a=2$ m/s${}^2$, $b=1.5$ m/s${}^2$, $T=1.4$ s, $s_0 = 2$~m,
$t_{\rm r}=2$ s, $l=5$ m, $d_0 = 1.2$~m, and $\sigma_0 = 2$. Note that a probably less realistic model with fewer 
parameters could be used as well.

We first show simulation results for $\sigma_1=0.6$. 
In our simulations, vehicles enter the empty road from $t=0$ on and the first pedestrian arrives after $t=500$ s. 
The road length is $L=1400$ m and the location of the crossing point $x_O=1200$ m.

Figure 1(c) shows the average pedestrian delay.
One can see that, when the arrival probability $p$ is small,
the average delay time essentially remains constant. However, when $p \gtrsim 0.05$, it begins to increase with growing 
values of $p$. Then, after reaching the maximum at $p \approx 0.32$, it goes down with a further increase of 
the arrival rate $r=p/dt$.

\begin{figure}
\begin{center}
\vspace*{-3mm}
\includegraphics[width=9cm]{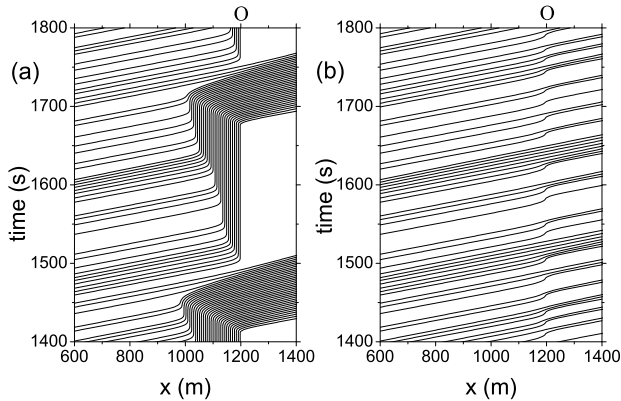}\\[-7mm]
\caption{(a) Representative space-over-time plot of vehicle trajectories for $\sigma_1=0.6$. Sometimes, vehicles are stopped by 
crossing pedestrians and form a queue. This suppresses the crossing of newly arriving pedestrians for a
long time period later on. (b) Representative space-over-time plot of vehicle trajectories for 
$\sigma_1=1.0$. Obviously, vehicles are decelerated, but not stopped by pedestrians.}
\end{center}
\end{figure}

Next, let us try to explain the change of the slope. For this, we show in Fig. 2(a) the typical structure
of the traffic situation obtained at $p=0.25$.
One can see that at the point O ($x=1200$~m), there is an alternating vehicle and pedestrian flow.
When pedestrians cross the road, the vehicles are stopped. When the stopped vehicle queue starts [this occurs when
no pedestrian arrives in the subsequent 20 time steps (corresponding to the crossing time $t_{\rm r} =2$ s)], 
the formed vehicle queue does not allow pedestrians to cross until a large gap occurs, again.

Figure 3(a) sketches the underlying mechanism. If pedestrians have stopped a vehicle at time $t_0$, the 
following vehicles queue up with a speed of \cite{control} 
\begin{equation}
 C = \left( \frac{\rho_{\rm jam}}{Q_{\rm arr}} - \frac{1}{v_0} \right)^{-1} \, ,  
\end{equation}
and more pedestrians can cross. $\rho_{\rm jam}$ is the jam density, 
$Q_{\rm arr}$ the arrival rate of vehicles and $v_0$ their desired velocity.
After a time interval $\Delta t_1$, i.e. 20 time steps after the last pedestrian has entered
the road, the first vehicle in the queue can accelerate again. The last vehicle in the queue reaches point O at
time $t_0 + \Delta t_1 + \Delta t_2$ with \cite{control}
\begin{equation}
\Delta t_2 = C \, \Delta t_1 \frac{1+ |c|/v_0}{|c| - C} \, ,
\end{equation} 
where $c = -1/(\rho_{\rm jam} T) \approx -15$ km/h is the characteristic jam resolution speed. 
Afterwards, pedestrians have the chance to find a gap in the vehicle flow, again.
\par\begin{figure}
\begin{center}
\hspace*{-0.5cm}\includegraphics[width=4.5cm]{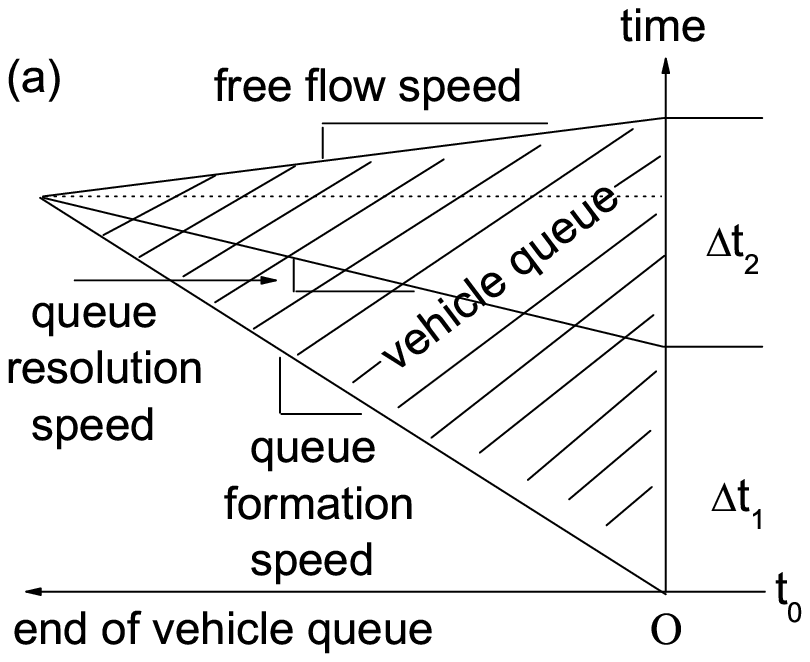} 
\includegraphics[width=4.5cm]{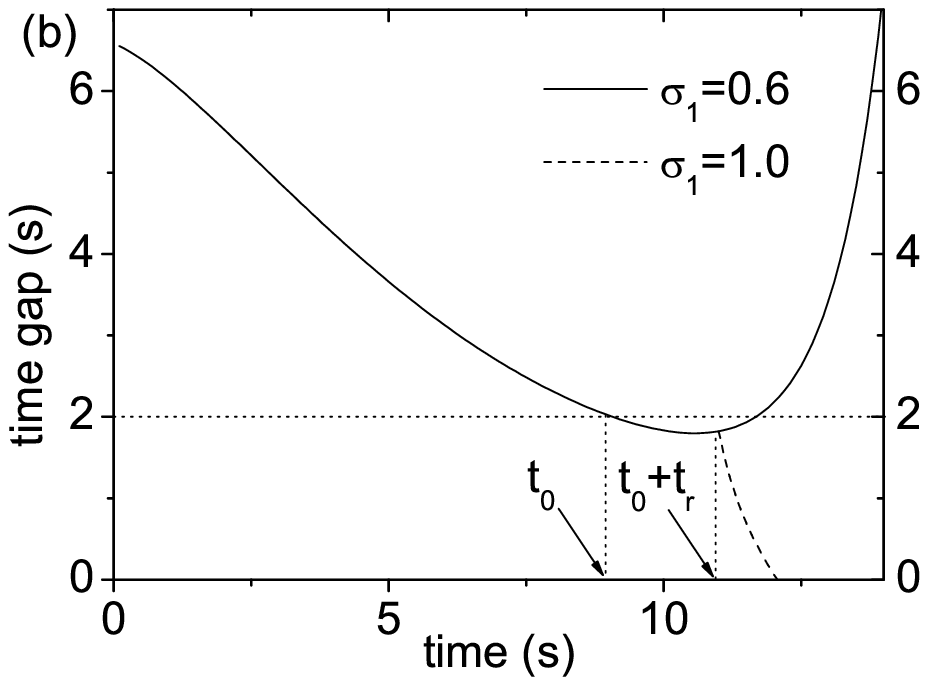}
\vspace*{-1cm}
\end{center}
\caption{(a) Sketch of the queue formation and queue resolution mechanism (see text), neglecting details of
acceleration and deceleration.
(b) Dynamically changing time gap $[d(t)-d_0]/v(t)$ of a vehicle if pedestrians
enter the street 100 meters ahead (see text). If $\sigma_1 t_{\rm r} \lesssim 2$, pedestrians
continue entering the street, which may stop the vehicle (solid growing curve). Otherwise, the crossing criterion
is violated after some time and the vehicle can accelerate (dashed falling curve).}
\end{figure}
During the period of $\Delta t_2$, the average number of arriving pedestrians is $p \, \Delta t_2$. 
We can assume that the waiting time of the last pedestrian is approximately zero, while it is approximately $\Delta t_2$ for the first one.
The average delay is $\Delta t_2/2$, and the overall expected delay of all waiting pedestrians amounts to
$p (\Delta t_2)^2/2$. 
As the delay to the pedestrians arriving during the time period $\Delta t_1$ is zero, the average delay is
\begin{equation}
  \frac{p (\Delta t_2)^2/2 + 0\cdot p\, \Delta t_1}{p(\Delta t_1+\Delta t_2)} = \frac{(\Delta t_2)^2}{2(\Delta t_1+\Delta t_2)} \, .
\end{equation}
With the increase of $p$, the probability $(1-p)^{20}$ that no pedestrian arrives in the 20 subsequent time steps 
decreases. This implies that the average value of $\Delta t_1$ goes up, which leads to an increase in the average delay.

Nevertheless, when $p$ is large, the stopped vehicle queue may
occupy the whole length of the road upstream of point O. 
In this case, $\Delta t_2$ will become a constant. Therefore, with a further increase of $p$, the average delay 
$\frac{(\Delta t_2)^2}{2(\Delta t_1+\Delta t_2)}$
decreases, as $\Delta t_1$ increases with $p$ \cite{footnote5}.

Let us now have a look at the average traffic flow of vehicles shown in Fig.~1(d).
One can see that, when $p<0.28$, the vehicle flow remains essentially constant. When $p>0.28$, it decreases with an increase of $p$.
This is because a stopped vehicle queue may occupy the whole street upstream of point O for a certain time 
period, so that the entry of further vehicles becomes temporarily impossible. In other words, the crossing pedestrian flow 
affects the street capacity for $p \ge 0.28$. 

We will now increase the safety coefficient of pedestrians to $\sigma_1 = 1.0$. This introduces an additional 
safety time gap for crossing the road and, thereby, reduces the impact on 
approaching vehicles [see Fig.~2(b) for $p=0.25$]. Figure~1(c) shows simulation results for 
$\sigma_1 = 0.6$ and $\sigma_1=1.0$. One can see that, compared with the result for
$\sigma_1 = 0.6$, the average delay time 
is slightly higher at small values of $p$, but it is considerably smaller for large arrival probabilities $p$.

In order to understand this qualitatively different result, we have studied 
the deceleration process of a freely moving vehicle
when a driver sees pedestrians entering the street $\Delta x = d = 100$~m 
ahead, with arrival probability $p=1$. In Fig.~3(b), the solid line shows the evolution of the
time gap $[d(t)-d_0]/v(t)$ with time $t$ in the case of $\sigma_1 = 0.6$. First,
the time gap decreases with time. However, after it reaches a minimum value, it increases
again. During the whole deceleration process, the time gap is larger than $\sigma_1 t_r$. This means
that pedestrians can always enter the road, and the vehicle may be stopped by pedestrians. However,
for $\sigma_1 = 1.0$, pedestrians will not enter the road after the time gap has decreased to 
$\sigma_1 t_{\rm r} = 2$~s at time $t_0$, as the safety criterion (\ref{safety}) becomes violated.
Nevertheless, the vehicle continues decelerating during the crossing time $t_{\rm r}=2$~s.
Afterwards, the vehicle accelerates again [dashed falling curve in Fig.~3(b)] and no pedestrian can
enter before the vehicle has passed point O. This means that pedestrians will never stop vehicles.
Consequently, no vehicle queue will form and
pedestrians will cross the street one by one or in small groups [see Fig.~2(b)]. 
Our simulations show that the transition between the continuous and oscillatory crossing
dynamics occurs at $\sigma_1 \approx 0.96$.

We have investigated 
the coupled pedestrian-vehicle delay problem with a simplified model. In contrast to our approach,
previous pedestrian delay models have neglected the influence of pedestrian crossing on vehicle dynamics. 
Our computer simulations were carried out for two different types of pedestrians to highlight the transition of
the system behavior at $\sigma_1\approx 0.96$:
(i) Aggressive pedestrians with a small safety coefficient $\sigma_1<0.96$ could force vehicles to stop
by successive crossing events, which produced alternating vehicle and pedestrian flows at high
pedestrian arrival rates $r = p/dt$. 
(ii) Careful pedestrians with a safety coefficient $\sigma_1 > 0.96$ did not stop vehicles and crossed the
street one by one or in small groups. Altogether, this mode was more efficient for pedestrians {\em and} cars,
as single pedestrian could not keep a growing vehicle queue
from going. 

As in panicking crowds \cite{panic}, impatience affects system performance in a negative way:
Waiting a bit longer (for a larger vehicle gap) implies smaller average delays
(``slower-is-faster effect''). Therefore, when the vehicle flow is not too large, a traffic light is not needed to allow
pedestrians to cross the street. It is rather required to terminate inefficient pedestrian
crossing while vehicle queues are building up \cite{footnote7}. Future investigations should clarify, 
whether the emergent oscillations in the density oscillator, ticking hour glass,
pedestrians streams, ant traffic, and collective motion of molecular motors
are also inefficient, as in the surprising example discussed here, or efficient, as expected. 

{\em Acknowledgements:} The authors thank for partial financial support by 
the Chinese National Natural Science Foundation (Grant No. 10404025 and 10272101), 
the Alexander von Humboldt Foundation, the German Research Foundation (DFG project He2789/7-1),
and the DAAD.


\begin{thebibliography}{10}
\bibitem{footnote1}
{Since the vehicles have a finite size, it is likely that the
front of a vehicle is downstream from point O. In this case,
$d$ may be regarded as negative and the vehicle blocks point
O. Thus, the pedestrian cannot traverse (see Fig.1(b)), which
is consistent with our above criterion, because $d/v$ is negative and
the condition (\ref{safety}) is not met.}
\bibitem{footnote3}
{Sometimes the stopped vehicles may extend upto the
entrance.}
\bibitem{footnote5}
In a certain range of $p$-values,
although the stopped vehicle queue may sometimes occupy the whole road upstream of point O, 
the average delay still increases with $p$ because the mechanism illustrated in Fig.~2(a) dominates.
\bibitem{footnote7}
The situation is different for long-lasting dense vehicle flows,
which cannot be crossed without a traffic light.
\end{thebibliography}
\end{document}